# Stability Limits and Surface Chemistry of Ag Nanoparticles in Non-Adsorbing Electrolytes Probed by Bragg Coherent Diffractive Imaging

Y. Liu[1], P. P. Lopes[1], W. Cha[1,2], R. Harder[2], J. Maser[2], E. Maxey[2], M. J. Highland[1], N. Markovic[1], S. Hruszkewycz[1], G. B. Stephenson[1]*, H. You[1]*, and A. Ulvestad[1].

[1]Materials Science Division, Argonne National Laboratory, Argonne, Illinois 60439, USA

[2]Advanced Photon Source, Argonne National Laboratory, Argonne, Illinois 60439, USA

*you@anl.gov, gbs@anl.gov

Surface chemistry is important across diverse fields such as corrosion[1–3] and nanostructure synthesis[4,5]. Unfortunately, many as-synthesized nanomaterials, including partially dealloyed nanoparticle catalysts for fuel cells[6–9], with highly active surfaces are not stable in their reactive environments, preventing widespread application. Thus, understanding instability by focusing on the structure-stability and defect-stability relationship at the nanoscale is crucial and will likely play an important role in meeting grand challenges[10]. To this end, recent advances in imaging nanostructure stability have come via both electron[11–16], x-ray[17–20], and other techniques such as atomic force microscopy[21], but tend to be limited to specific sample environments and/or two-dimensional images. Here, we report investigations into the defect-stability relationship of silver nanoparticles to voltage-induced electrochemical dissolution imaged *in-situ* in three-dimensional (3D) detail by Bragg Coherent Diffractive Imaging (BCDI). We first determine the average dissolution kinetics by Stationary Probe Rotating Disk Electrode (SPRDE) in combination with inductively coupled plasma mass spectrometry (ICP-MS), which allows real-time *in-situ* measurement of Ag$^+$ ions formation and the corresponding electrochemical current. We then observe the dissolution and

**redeposition processes in 3D with BCDI in single nanocrystals, providing unique insight about the role of surface strain, defects, and their coupling to the dissolution chemistry. The methods developed and the knowledge gained go well beyond a "simple" silver electrochemistry and are applicable to all electrocatalytic reactions where functional links between activity and stability are controlled by structure and defect dynamics.**

Elucidating the role of structure and defects in electrochemistry is a topic of broad based scientific as well as technological importance; in particular across diverse fields such as electrocatalysis[22], synthesis[23] and corrosion[24,25]. Not surprisingly, then, the last decade has witnessed substantial advances in our understanding of the functional links between the surface defect density and the reactivity-stability of electrochemical interfaces[]. These ongoing developments have been driven primarily by the emergence of structural probes, most notably scanning tunneling microscopy (STM) and surface x-ray scattering (SXRS), which are capable of visualizing ordered surface structures and surface defects such as surface adatoms, steps and kinks. In combination with a "simple" rotating disk electrode (RDE) coupled with plasma mass spectrometry (ICP-MS), these two probes have offered an ability to correlate surface chemical reactivity and stability with atomic-level surface structure.

One important, yet experimentally unexplored, question concerns the role of surface strain and extended defects such as dislocations which, based on recent computational approaches, may be key in governing properties at electrochemical interfaces[]. Strain is the physical concept used to describe the deviations from the ideal crystal structure and it is accepted that it arises from chemical, electrical, magnetic and other forces in the crystal[26,27]. Because of the important role of their surfaces, strain in nanoparticles is enhanced relative to their bulk counterparts, thus opening significant opportunities for establishing relationships between the nanocrystal strain and nanotechnology functionality. While STM and traditional SXRS still constitute important structural probes, they have inherent limitations for visualizing surface strain and extended defects; and, in turn, how these affect interior dynamics. One the other hand, Bragg coherent diffractive imaging (BCDI) utilizing hard x-rays has garnered considerable attention for *in-situ* imaging of the three-dimensional (3D) distribution of strain and dislocations[28–30]. Thus far, BCDI is considered to be a "standard tool" for imaging strain in isolated nanoparticles. Application of the same method in electrochemistry is, however, still in its infancy. Considering the importance of surface defects and surface strain in controlling physicochemical properties of metal-solution interfaces, encompassing those of catalytic and corrosion relevance, exploiting BCDI in electrochemical environments is of broad-based fundamental and practical significance.

This fact triggers the intriguing question of what system to choose in order to be able to exploit the full power of BCDI in monitoring the role of strain and defects in both the dissolution of surface atoms (corrosion) as well as the evolution of new surfaces during re-deposition of the dissolved hydrated cations. As mentioned above, although corrosion and deposition processes are central to many important applications[], the role of surface strain and extended defects is not well understood. If established, however, it may open new avenues not only for understanding corrosion at the atomic level but, in addition, it may offer novel synthesis methods for tailoring activity and stability of materials by an optimal density of strained surface atoms.

Here, we use *in-situ* BCDI to explore an intimate relationship between the 3D strain distribution and dissolution/redeposition of silver nanocrystals (Ag-NP with an average size of ca. 300-500 nm) in perchloric acid solutes. We chose silver electrochemistry because it has served as a model system for both understanding electrochemical dissolution (corrosion) processes as well for elucidation key parameters that control the concomitant kinetics of silver deposition[31–34]. The real time kinetics of silver dissolution/redeposition are established using Stationary Probe Rotating Disk Electrode (SPRDE) in combination with ICP-MS. Taken together with BCDI, by visualizing the local 3D morphology and dislocation distributions during dissolution and redeposition at the single nanoparticle level it was possible to find interesting relationships between the surface strain, defects, and the mechanism of silver desolution/redeposition at electrochemical interfaces. The methods developed and the knowledge gained go well beyond a "simple" silver electrochemistry and are applicable to all electrocatalytic reactions where functional links between activity and stability are controlled by the dynamics of defect formation and surface strain.

Silver nanoparticles were synthesized on polished glassy carbon disks by annealing of sputtered silver films. The thermal annealing leads to isolated nanoparticles that are nearly perfect single crystals (see **Methods**, **Figure 1**) with different orientations relative to the substrate. ICP-MS measurements were carried out in 0.1 M $HClO_4$.

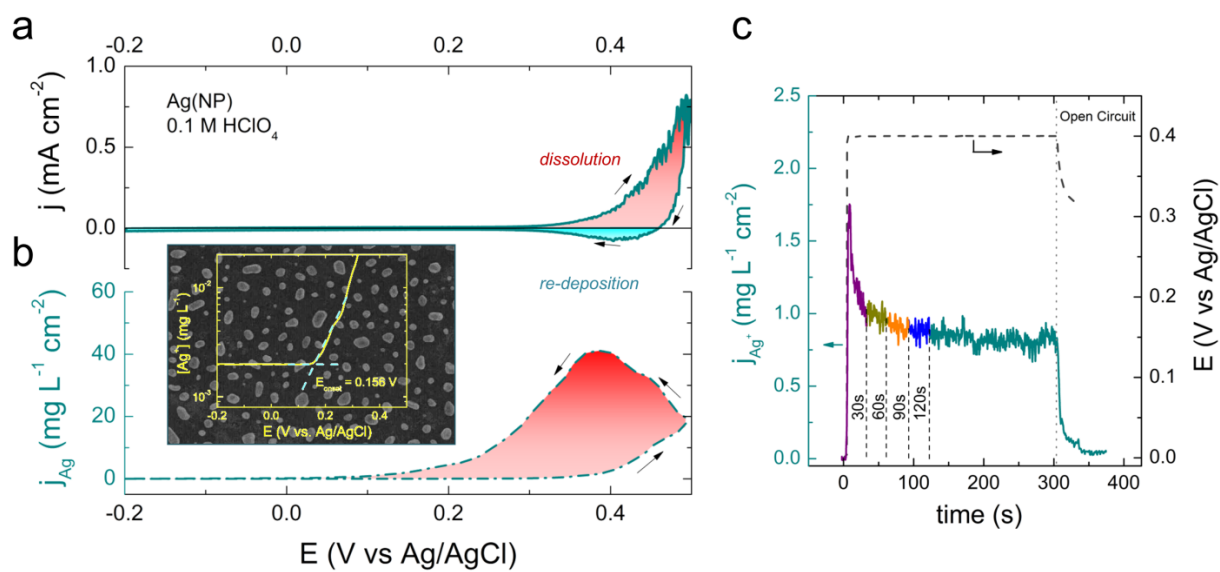

**Figure 1. Overall dissolution kinetics observed through ICP-MS**. **(a)** Cyclic voltammetry and **(b)** simultaneous *in-situ* monitoring of $Ag^+$ ions production with SPRDE-ICP-MS during electrochemical dissolution of Ag nanoparticles (SEM inset in **b**, 25 um

by 25 um) at 50mVs$^{-1}$ in 0.1M HClO$_4$. Early signs of Ag dissolution can be observed at 0.156V in clean electrolytes (inset in **b**). **(c)** Potential step to 0.4 V and simultaneous dissolution of Ag$^+$ over time for total 300 seconds, followed by opening the circuit. Initial 30, 60, 90 and 120 seconds intervals highlighted as the dissolution rate decreases substantially within this period.

**Results and Discussion**

**Kinetics of silver dissolution and re-deposition**. **Figure 1** reveals the overall dissolution kinetics determined by SPRDE coupled to ICP-MS. As shown in **Figure 1a**, the cyclic voltammetry profile for Ag(NP) shows dissolution during the positive sweep (anodic scan) and dissolution and redeposition during the negative sweep (cathodic scan). Concomitantly, in **Figure 1b** it is shown that the formation of Ag$^+$ begins in the anodic scan and continues during the cathodic sweep as a consequence of the fast sweep rate (**Supplementary Fig. 1**). While the voltammetry data suggests the onset of dissolution at around 0.3V vs Ag/AgCl, the ICP-MS data (inset in **Fig. 1b**) determines the onset potential at 0.156V+/- 0.010V. This is in excellent agreement with thermodynamic predictions from the Nernst equation given the initial concentration of 2 ppb of silver ions ([Ag$^+$]~20 nM). This indicates that the kinetics of Ag dissolution are so fast that the current vs. potential profile is dictated mainly by limited transport of dissolved silver from the diffusion layer. It is worth noting that small changes in the background concentration of silver will not significantly affect the main voltammetry profile, and only changes in the onset of dissolution as measured by ICP-MS are observed (**Supplementary Fig. 2**). Nonetheless, for every consecutive voltammetry more silver is being removed than re-deposited, which will eventually lead to smaller total surface area, as only a limited number of nanoparticles are present in the sample. This effect is depicted in **Figure 1c**, where a potential step to 0.4V triggers Ag dissolution and Ag$^+$ formation rate decreases with time, most notably in the first 60 seconds. While ICP-MS clearly resolves the electrochemical kinetics of silver dissolution, it is blind to role of surface morphology and defects in the dissolution process, which we image in single nanocrystals using BCDI.

**3D Imaging of the atomic displacement field: "surface" dynamics during dissolution.** In BCDI, the 3D intensity distributions around an isolated Bragg peak are collected by slightly rotating the crystal with respect to the incident x-ray beam. The 3D intensity distribution is then Fourier transformed into a real space image via phase retrieval algorithms[35–38]. The real space image is complex, with the amplitude corresponding to the diffracting or Bragg electron density and the phase corresponding to a projection of the atomic displacement field onto the chosen scattering vector[30,39,40]. **Supplementary Fig. 3** shows a schematic of the *in-situ* three-electrode electrochemical cell (see **Methods**). This cell uses a "thin layer" of electrolyte (thickness ~300 microns) to reduce x-ray attenuation and as such the kinetics will differ from the SPRDE measurement. Real space images of the Ag nanocrystals were reconstructed with between 20 to 30 nm spatial resolution as determined by the phase retrieval transfer function[41,42] (see **Supplementary Fig. 4**). For further details, please see the **Methods**. Based on the dissolution thermodynamics (**Fig. 1, Supplementary Fig. 1**), we began our dissolution experiments by applying 0.4 V relative to the Ag/AgCl reference electrode.

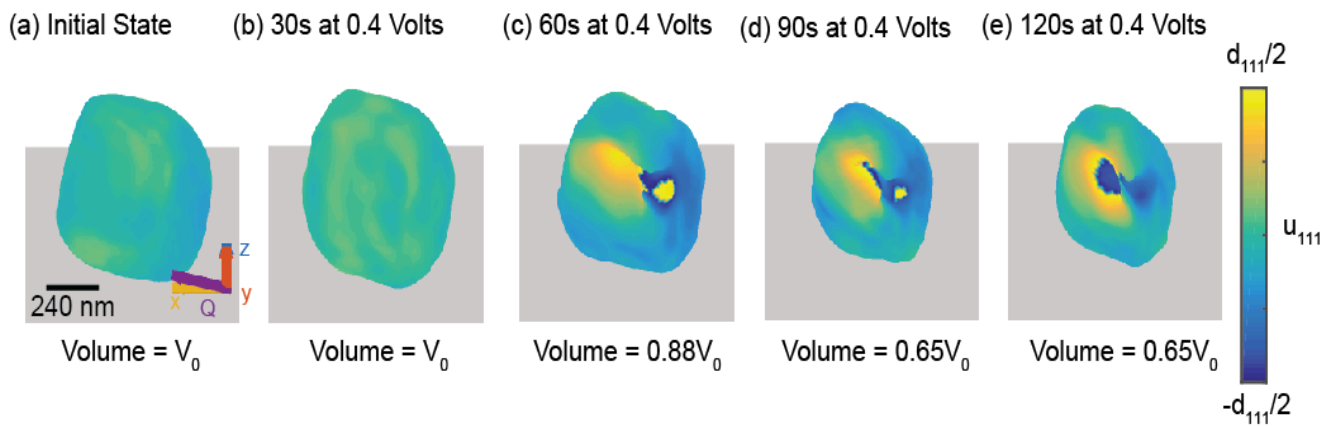

**Figure 2**. 3D view of single particle dissolution and surface displacement

**evolution**. The scattering vector (Q) and the lab frame axes (x, y, z) are shown. The remaining volume as a fraction of the initial volume is indicated. The "surface" displacement (colormap) should be considered an average over the outermost 20-30 nm of the crystal. **(a)** The as-synthesized particle shape (isosurface) and the displacement field (colormap projected onto the isosurface). The colorbar applies to all images. **(b)** After 30s at 0.4 V. **(c)** After 60s at 0.4 V. **(d)** After 90s at 0.4 V. **(e)** After 120s at 0.4 V. Voltage pulsing causes both a decrease in particle size and a dramatic increase in the surface atoms displacement from their equilibrium values. Although the particle decreases in size at all surfaces, there is more relative dissolution from the corner with the strong displacement field variation in **d-e**.

To explore the connection between surface strain, defects, and dissolution/redeposition dynamics, we chose to image an initially defect-free silver nanocrystal with a non-uniform strain state. The particle is approximately 530x430x480 nm$^3$ (shown by the isosurface), with regions of primarily positive surface displacement (colormap projected onto the isosurface), especially near a particular corner (**Figure 2a**). We note that the "surface" displacement field refers to an average over a 20-30 nm region (the spatial resolution) at the boundary of the particle. For these measurements, the voltage was increased from open circuit voltage (OCV) to 0.4 V for the indicated time and then returned to OCV. The primary change in the particle after the first voltage pulse is in the surface displacement of the atoms from their equilibrium positions, which increases across the surface (**Figure 2b**). A further 30s pulse causes a change in the particle's volume and the surface displacement field magnitudes to approach half of a silver [111] lattice spacing (**Figure 2c**). The particle then decreases in size more rapidly near the region where the strong displacement fields are located (**Figures 2d-e)**. These strong displacement fields are consistent with the displacement fields generated by dislocations[43,44]. **Supplementary Fig. 5** shows an additional 3D view of the dissolution process for the same time states. By observing the surface displacement during the dissolution process, we conclude that initial surface strain and surface defects likely play an important role in both the onset of dissolution and continuing dissolution kinetics. Of course, there is always both dissolution and redeposition when returning the voltage to OCV. In this case, however, dissolution dominates redeposition as the particle size decreases during the voltage pulsing. We discuss an example where significant redeposition was observed later in the text. While **Figure 2** shows a surface view of the dissolution process, BCDI reveals dynamics throughout the particle volume, allowing insight into the surface-interior coupling throughout the dissolution process. We now discuss the changes induced in the particle interior due to this coupling.

**3D Imaging of the atomic displacement field: interior dynamics during dissolution**

Three cross-sections through the particle (see **Supplementary Fig. 6** for their spatial location in an isosurface view) were chosen to show the dissolution-induced changes in the particle interior. The colormap is the displacement of the atoms from their equilibrium positions, while the shape corresponds to the Bragg electron density.

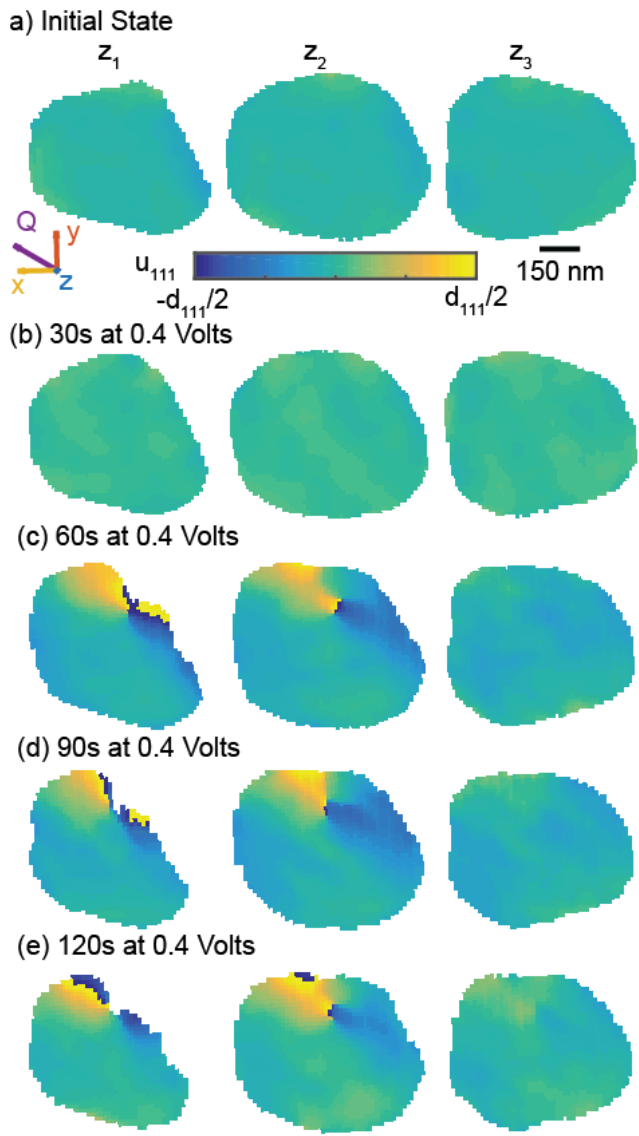

**Figure 3. Three 2D cross-sections of dissolution-induced displacement field evolution due to surface-bulk coupling.** The scattering vector (Q) and the lab frame axes (x, y, z) are shown. Color scale applies to all images. Cross-sections in the particle: **(a)** The as-synthesized particle. **(b)** After 30s at 0.4 V. **(c)** After 60s at 0.4 V. **(d)** After 90s at 0.4 V. **(e)** After 120s at 0.4 V. The particle initially shows the largest displacements near its surface, in particular near the upper right corner in slice $z_1$. This initial state has an effect on the dissolution onset, which occurs initially near this corner.

Large displacement field changes both near the surface and in the bulk are induced during dissolution. While dissolution happens at all particle surfaces, more relative dissolution occurs close to the strong displacement field variations.

The initial interior displacement field is low in magnitude (**Figure 3a**) and the largest tensile strains ($\partial_{x_{111}} u_{111}$) exist near the upper right corner of the leftmost cross-section ($z_1$) (**Supplementary Fig. 7**). We can now visualize how the initial surface strain affects the surface and interior dissolution during applied voltage. The first 30s voltage pulse causes the magnitude of the displacement field to increase throughout the particle (**Figure 3b**). The second 30s pulse leads to the development of strong displacement field variations, including displacements up to $\pm d_{111}/2$, in multiple cross-sections of the particle both at the surface and in the interior (**Figure 3c).** Displacement field vortices, which are circular paths along which the displacement field varies from $-d_{111}/2$ to $+d_{111}/2$, that cannot be removed by a global phase offset are consistent with the displacement fields generated by dislocations[43,45]. In addition, the spatial locations at which dislocations terminate on the particle surface correspond to a region where significant dissolution has occurred (slice $z_1$, **Figure 3c**). The preferential dissolution of the Ag nanocrystal corner is consistent with two observed phenomena: preferential dissolution rates of different facets[10], and enhanced dissolution of locally strained regions[24]. The central cross-section shows that the dislocation displacement field extends approximately 150 nm into the crystal and leads us to conclude that the surface dissolution dynamics significantly affect the interior crystal response during dissolution. **Figures 3d-e** show the displacement field evolution during further voltage pulsing.

With the ability to resolve the entire 3D morphology and displacement field information, BCDI has shown there is a connection between initial surface strain, surface and interior defects, and relative dissolution rates. The initial tensile strain at the particle corner corresponds both to the point at which significant dissolution first occurs and the dislocation location. By resolving the interior structure, we see that the surface influences the bulk response and that the surface dislocations extend significantly into the particle. It is known that different defect types affect the dissolution behavior in different ways[46], and thus it is useful to identify the nature and type of the dislocation in the particle. In order to identify the defect, we map the dislocation line in 3D using a gradient-based method (see **Methods**).

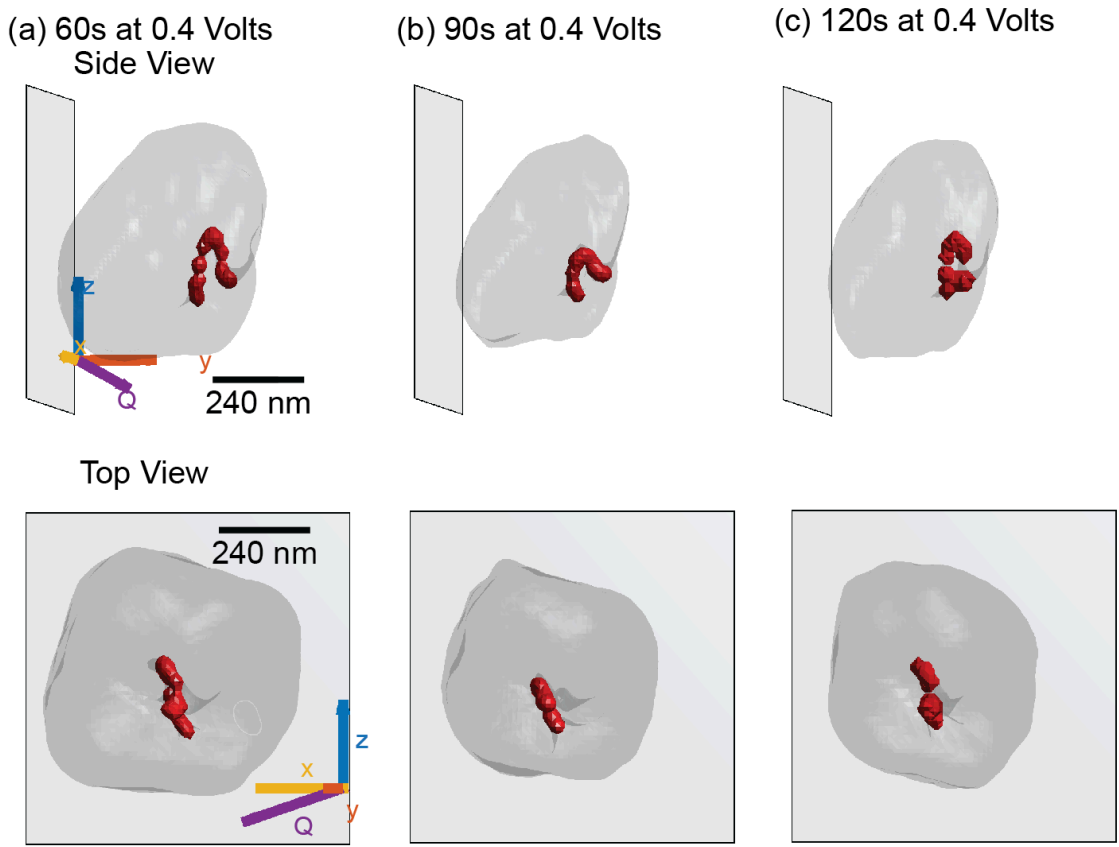

**Figure 4**. **3D dislocation line mapping, identification, and evolution due to dissolution**. The dislocation line is traced out in red while the particle shape is shown by the semitransparent black isosurface. Two different views are shown. The lab frame coordinates (x, y, z) and the scattering vector (Q) are shown. Only the states after

dislocation creation are shown: **(a)** After 60s at 0.4 V. **(b)** After 90s at 0.4 V. **(c)** After 120s at 0.4 V. The dislocation line exhibits a horseshoe shape and lies in a single plane, consistent with a mixed dislocation line. It terminates at two points on the particle surface.

**3D Imaging of the displacement field: identifying the dislocation**

The particle shape is shown as a black isosurface with the 3D dislocation line traced in red for the first state after dislocation nucleation (**Figure 4a).** The side view shows that the dislocation line follows a horseshoe-like path in which the line originates near the particle boundary, continues into the particle before making two 90 degree turns and terminates at the particle surface at a different location approximately 80 nm away. While the side view shows the loop, the top view shows that the loop lies essentially in a single plane. This type of dislocation line distribution is consistent with a mixed dislocation loop that makes 90 degree turns by transitioning from an edge to a screw dislocation (or vice versa)[47]. Without the 3D information provided by BCDI, the observation of a loop structure and subsequent defect identification would be impossible. For completeness, we also show the dislocation line after further voltage pulsing (**Figures 4b-c)**.

We observed that an initially surface strained, defect-free nanocrystal undergoes significant changes in both the Bragg electron density and the displacement field in response to the applied voltage (**Figures 2-4)**. The initial surface strain influences the spatial location of dissolution onset, which also induces dislocation formation. Dislocations are identified as mixed type from the 3D geometry of the dislocation line. Enhanced relative dissolution is seen at the termination points of the dislocation line on the particle surface. An example of selective dissolution in a particle with an initial dislocation is shown in **Supplementary Fig. 8**. The total amount of dissolution for each voltage pulse is not constant for any of the particles that were imaged, which is consistent with the ICP-MS data (**Fig. 1c**). Decreasing of the total dissolution amount for a given voltage pulse is due to the particle surface area decrease and that the most unstable crystal regions dissolve first, leaving behind more stable areas.

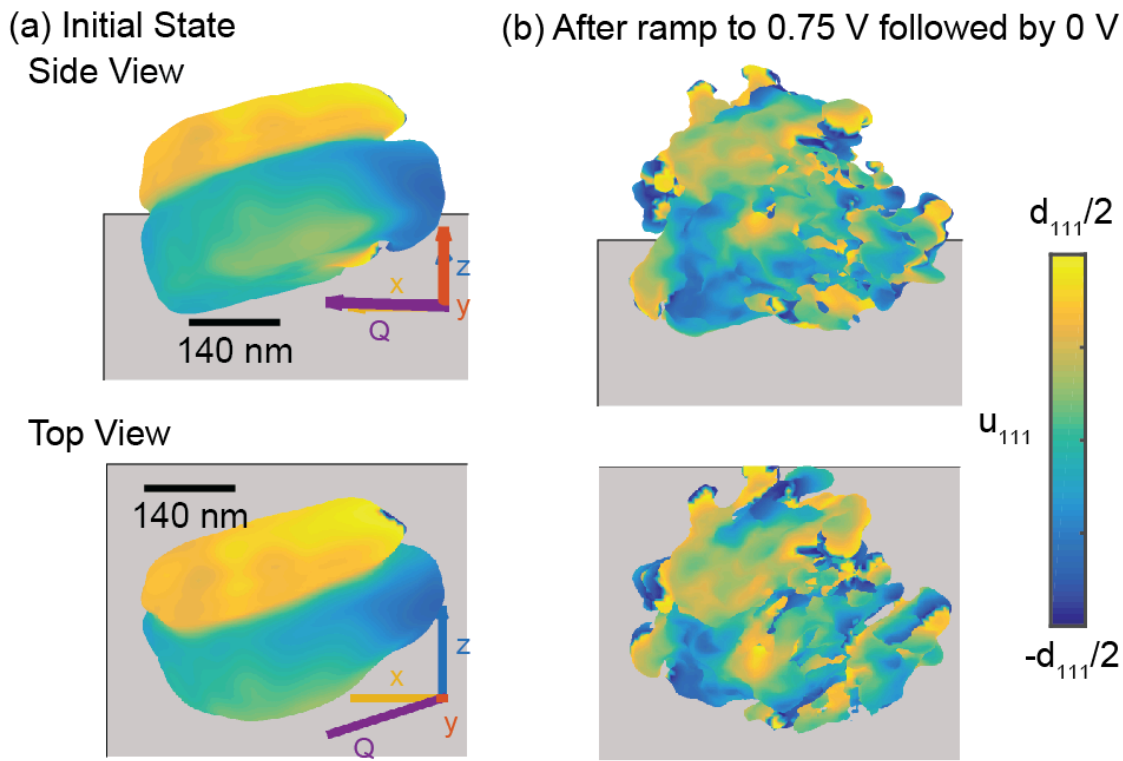

**Figure 5**. **3D view of redeposition after dissolution to link the initial and final nanocrystal morphologies**. Side and top views of the particle before and after the electrochemical procedure. The scattering vector (Q) and the lab frame axes (x, y, z) are shown. Colorscale applies to all images. **(a)** The as-synthesized particle shape (isosurface) and displacement field (colormap projected onto the isosurface). **(b)** After the voltage ramp to 0.75 V followed by the application of 0 V. Significant dissolution occurred during the voltage ramp followed by significant redeposition in registry with the original crystal lattice when the voltage is set to 0 V. Silver redeposition in the original crystal lattice orientation shows that the redeposited state is linked to the initial state.

**3D Imaging of the displacement field: a "surface" view during redeposition**

The high local concentration of Ag ions near the electrode surface offers an opportunity to study Ag deposition following its dissolution. Redeposition is a process commonly encountered in electrocatalyst employment and in the corrosion of structural materials as their local environments change. Redeposition can be produced in this system, for example, by setting the voltage to zero instead of returning to OCV. The deposition of dissolved Ag ions is well illustrated on the cyclic voltammogram, where the reduction current from Ag deposition on its cathodic scan is evident in **Fig. 1**. To study this process, we chose a different particle from a nominally identical sample that was previously unexposed to voltage biasing. To induce significant dissolution followed by redeposition, the voltage was ramped from 0 V to 0.75 V in 0.01 V increments with a 1s hold at each voltage followed by returning the voltage to 0 V. Two views (side and top) of the initial particle state are shown (**Figure 5a**). The isosurface shows the Bragg electron density while the color map shows the displacement field projected onto the isosurface. **Figure 5b** shows the same two views after the voltage cycle. Note that BCDI is exclusively sensitive to Bragg electron density[48] and as such sees the silver ions that redeposit in registry with the orientation of the original {111} crystal lattice planes. Thus, we observe that there is significant redeposition at exactly the same lattice orientation as the original crystal, though the morphology of the redeposited material varies greatly from that of the original state. The redeposition of atoms at the same lattice orientation shows that the end state after dissolution significantly influences the redeposition location. An example of an additional particle that underwent dissolution followed by redeposition is shown in **Supplementary Fig. 9**.

**Conclusions**

We studied electrochemical dissolution and redeposition of silver nanoparticles on average using SPRDE coupled with ICP-MS and at the individual nanocrystal level using BCDI. We observed inhomogeneous dissolution of the Ag nanocrystals and a connection between both surface strain and defects in controlling relative dissolution rates. Using the 3D displacement field information, we observed a horseshoe-like dislocation loop structure, consistent with a mixed dislocation, which nucleated near the region of most rapid dissolution and extended significantly into the crystal interior. This demonstrates a coupling between the surface and interior during dissolution. By controlling the potential, we induced and imaged significant redeposition. The redeposition of atoms in registry with the original crystal lattice shows that the dissolution end state influences the redeposition location. Our results and methods are applicable to all electrocatalytic reactions where functional links between activity and stability need to be understood. More generally, our methods are useful in any system in which the dynamics are influenced by defect formation and surface strain.

**Acknowledgements:**

This research used resources of the Advanced Photon Source, a U.S. Department of Energy (DOE) Office of Science User Facility operated for the DOE Office of Science by Argonne National Laboratory under Contract No. DE-AC02-06CH11357. Y.L., H.Y., S.O.H., and G.B.S. were supported by U.S. DOE, Basic Energy Sciences, Materials Sciences and Engineering Division. We acknowledge the Center for Electron Microscopy at Argonne National Laboratory.

**Author Contributions:**

P. L. and N. M. designed, performed, and analyzed the results of the ICP-MS experiment. Y. L., H. Y., and A. U. designed the BCDI experiment. Y. L., W. C., R. H., J. M., E. M., S. O. H., G. B. S., H. Y., and A. U. performed the BCDI measurement. Y. L., A.

U., and M. J. H. synthesized the Ag nanoparticles. All authors interpreted the results and contributed to writing the manuscript.

**Supplemental Material:**

Ag Nanoparticle Synthesis. Ag nanoparticles were formed via a dewetting procedure. Glassy carbon rods (6 mm diameter x 30 mm long), purchased from Alfa Aesar, were cut into 5mm long cylinders and polished with 3 micron-grit polishing paper on one side. The disks were rinsed with 18 MΩ water and dried in a nitrogen stream before transferred to an rf-sputtering system for deposition. A silver film of 25 nm thick was deposited on the polished side of the disc at a rate of 1.5 Å/s. The film, which had been stored in the ambient environment, was annealed at 850 $^0$C for two hours to form the Ag particles via film coalescence. The samples were housed in a quartz tube sealed on both ends using silicone plugs, which allowed gas flow. Throughout the annealing, the tube was filled with 3% $H_2$ in Ar gas. **Fig. 1** inset shows a sample SEM of the as-prepared sample.

ICP-MS. The cyclic voltammetry of the Ag electrode was done in a deoxygenated 0.1 M perchloric acid at a sweep rate of 0.05 V/s. The occurrence of an increasing positive current at 0.31 V in the positive potential scan indicates the onset of Ag dissolution. The dissolution onset is consistent with the Ag+/Ag equilibrium potential at 0.306 V derived from its standard redox potential for an electrolyte containing $1x10^{-5}$ M Ag ion. On the reverse scan, dissolved Ag ions begin depositing as the potential goes below 0.46 V. The deposition potential is 0.15 V more positive than the dissolution onset, which is a reflection of the Ag ion concentration increase near the electrode surface. The concentration approximately increases from $10^{-5}$ M to $3 x 10^{-3}$ M. The Ag ion concentration near the electrode surfaces is a balance of the Ag dissolution rate and the Ag ion diffusion, which are dependent respectively on the dissolution potential and the electrochemical cell geometry. The formation of a reduction peak centered at 0.4 is evidence of the depletion of Ag ions near the electrode surface. The reduction current below -0.5 V is attributed to hydrogen evolution from Ag surface.

3-electrode Electrochemical Cell. **Supplementary Fig. 3** shows a schematic of the x-ray compatible electrochemical cell used in this work. The cell body was made from KEL-F. . The cell uses a silver/silver-chloride reference electrode, a platinum wire as the counter electrode, a gold wire as the electrical contact to the working electrode (a glassy carbon disk), and 0.1 M perchloric acid as the electrolyte. Prior to the experiment, the cell was cleaned in nitric acid and then washed with 18 MΩ water. The silver-silver chloride reference electrode was from World precision Instruments, Inc. High-purity Au and Pt wires from Alfa Aesar were used. The electrolyte of 0.1 M perchloric acid was prepared by diluting the ultrapure acid from EMD with 18 MΩ water. Glassy carbon discs make contact with a gold wire (working electrode). An O-ring and an x-ray transparent membrane (polyethelyne) are used to hold the electrolyte.

Bragg Coherent Diffractive Imaging experiment details. Experiments were performed at Sector 34-ID-C of the Advanced Photon Source at Argonne National Laboratory. A double crystal monochromator was used to select E=8.919 keV x-rays with 1 eV bandwidth and longitudinal coherence length of 0.7 $\mu m$. A set of Kirkpatrick-Baez mirrors was used to focus the beam to $1\times 1\ \mu m^2$ (vertical x horizontal). The rocking curve around the Ag (111) Bragg peak was collected by recording 2D coherent diffraction patterns with an x-ray sensitive area detector (Medipix2/Timepix, 256x256 pixels, each pixel 55µm x 55µm). The Bragg angle for this experiment was $2\theta = 33°$. It was placed a distance of 0.8 m away from the sample and an evacuated flight tube was inserted between the sample and the camera. A total of 51 patterns were collected over an angular range of $\Delta\theta = \pm 0.2°$ for each 3D rocking scan. Each scan takes a total of 3 to 4 minutes to complete.

Phase retrieval. The phase retrieval code is adapted from published work [27,49]. The hybrid input-output[38,50] and error reduction algorithms were used for all reconstructions. 1050 total iterations, consisting of alternating 40 iterations of the hybrid input-output algorithm with 10 iterations of the error reduction algorithm, were run for all diffraction patterns. This process was repeated five times for each reconstruction with five different random starts. The best reconstruction from the series of five random starts (quantified by the smallest sharpness metric) was then used in conjunction with another random phase start as a seed for another 10 reconstructions with yet another set of random starts. The sharpness metric is the sum of the absolute value of the reconstruction raised to the 4$^{th}$ power. Five generations were used in this guided algorithm[51]. The final resolution of 20 to 30 nm was computed via the phase retrieval transfer function, an example of which is shown as **Supplementary Fig. 4**.

Identifying phase vortices using displacement field gradients

The discrete gradient is calculated from the displacement field:

$$\nabla u = \left(\frac{\partial u}{\partial x}, \frac{\partial u}{\partial y}, \frac{\partial u}{\partial z}\right)$$

for the complete range of phase offsets: $u \rightarrow ue^{i\phi}$ with $\phi: 0 \rightarrow 2\pi$. Only gradient singularities that are identified for all phase offsets are included as part of the dislocation line.

**Supporting Figures.**

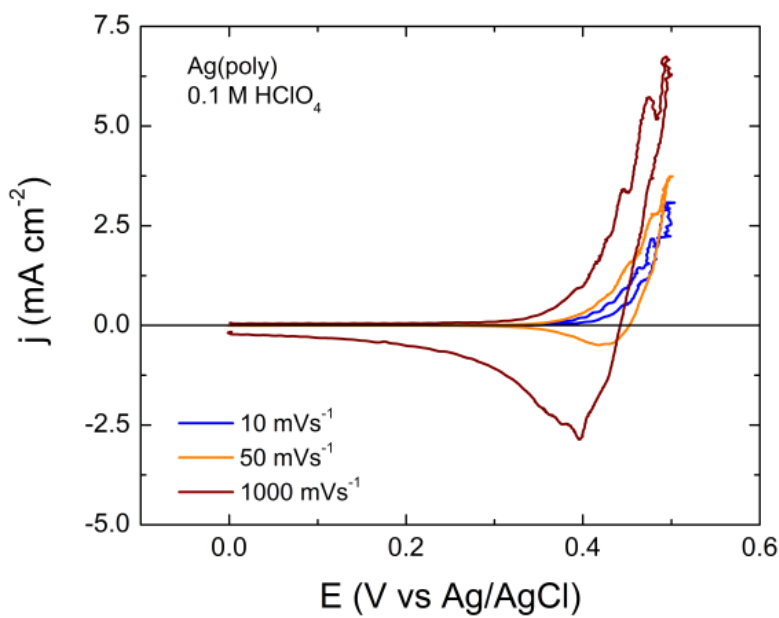

Figure S1 – Cyclic voltammetry of Ag(NP) at distinct sweep rates (10, 50 and 1000mV s$^{-1}$) to demonstrate dissolution and redeposition processes as indicated by positive or negative currents.

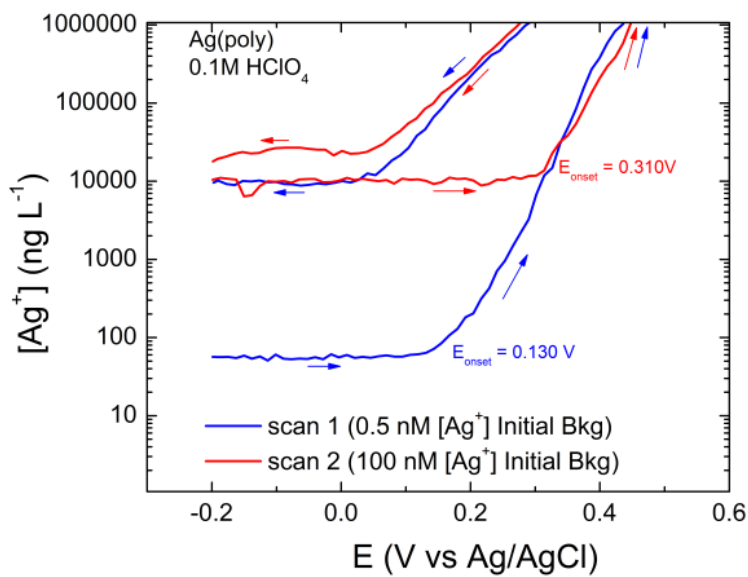

Figure S2 – In situ concentration of Ag ions produced from Ag(poly) during cyclic voltammetry experiment showing first and second scans and changes in apparent onset potential for silver dissolution with changes in background silver concentration.

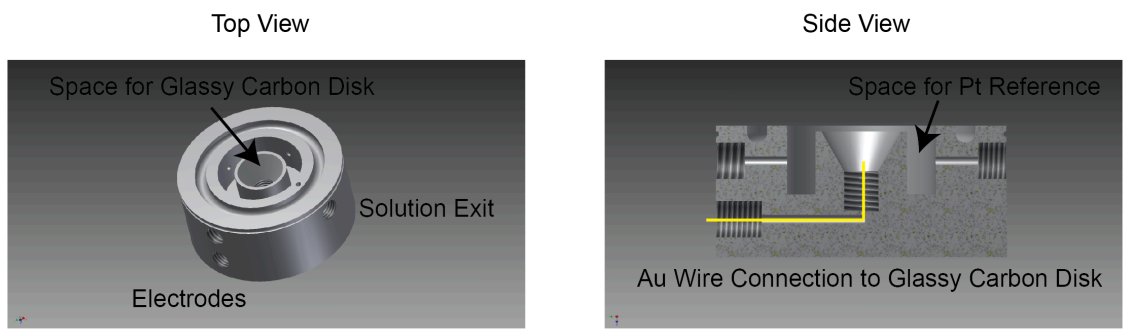

Figure S3. Two views of the in-situ electrochemical cell. Further details are given in the Methods.

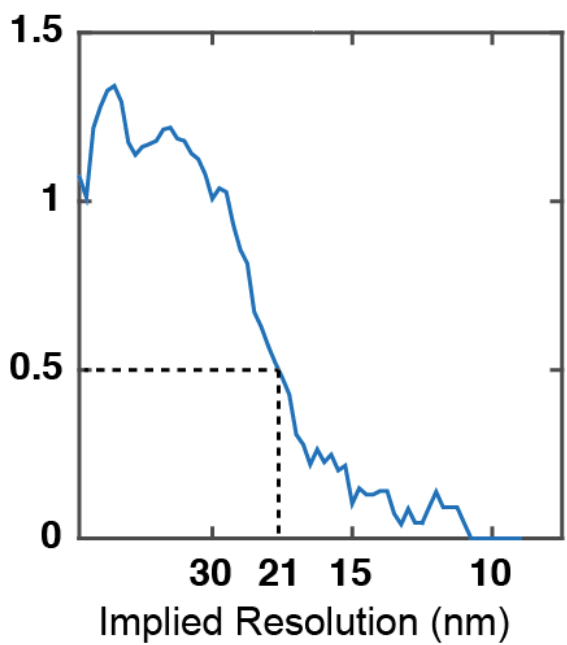

Figure S4. Phase retrieval transfer function for the reconstructions discussed in **Figures 2-4**. A conservative cutoff of 0.5 is used to determine the resolution as approximately 21 nm.

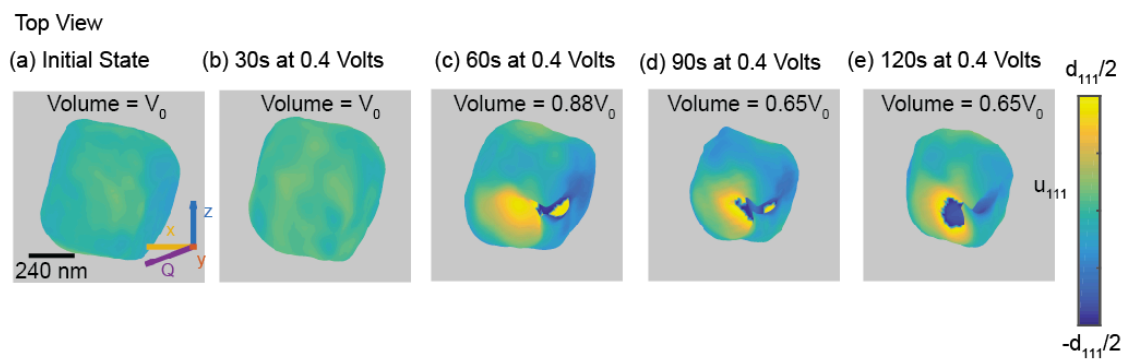

Figure S5. An additional view of the particle discussed in **Figure 2**. The scattering vector and the lab frame axes are shown. In each pane, the volume of the particle at each time state is shown, expressed as a fraction of the volume of the initial state $V_0$. **(a)** The as-synthesized particle shape and displacement field (colormap projected onto the surface). **(b)** After 30s at 0.4 V. **(c)** After 60s at 0.4 V. **(d)** After 90s at 0.4 V. **(e)** After 120s at 0.4 V.

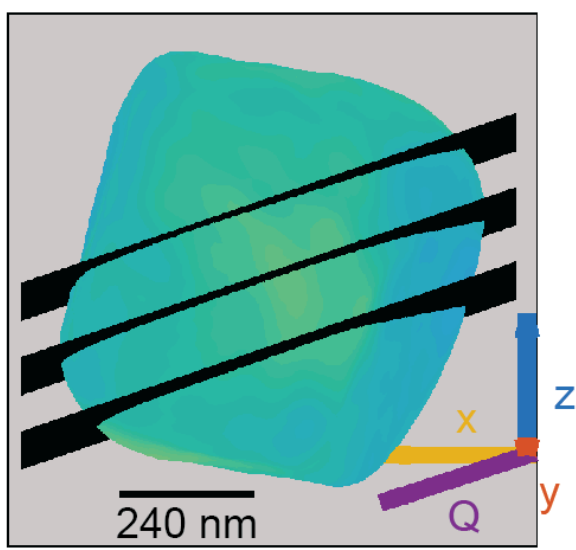

Figure S6. Cross-section locations in the first particle discussed in **Figures 2-4**. The top view is shown.

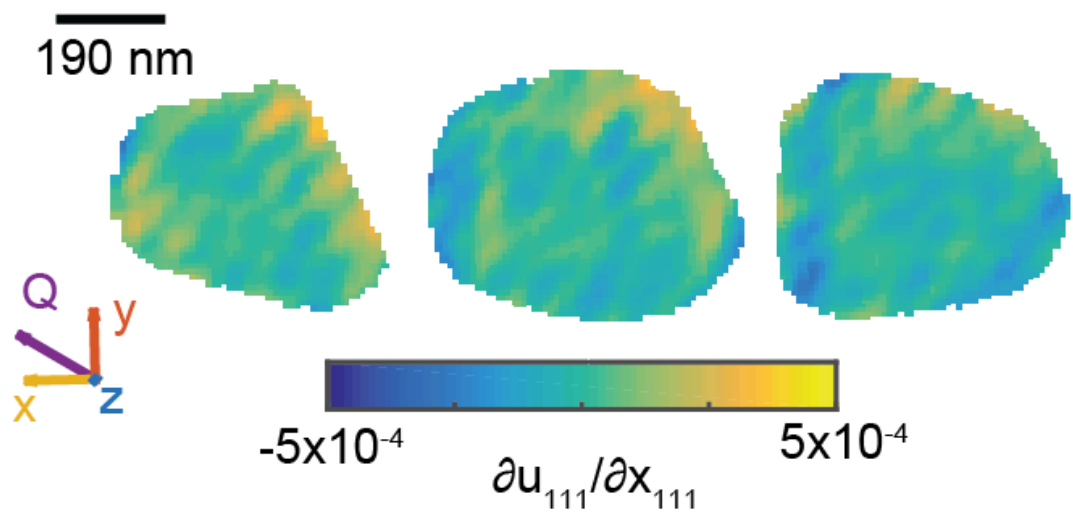

Figure S7. Compressive/tensile strain cross-sections for the as-synthesized state. The particle is the same as discussed in **Figures 2-4**.

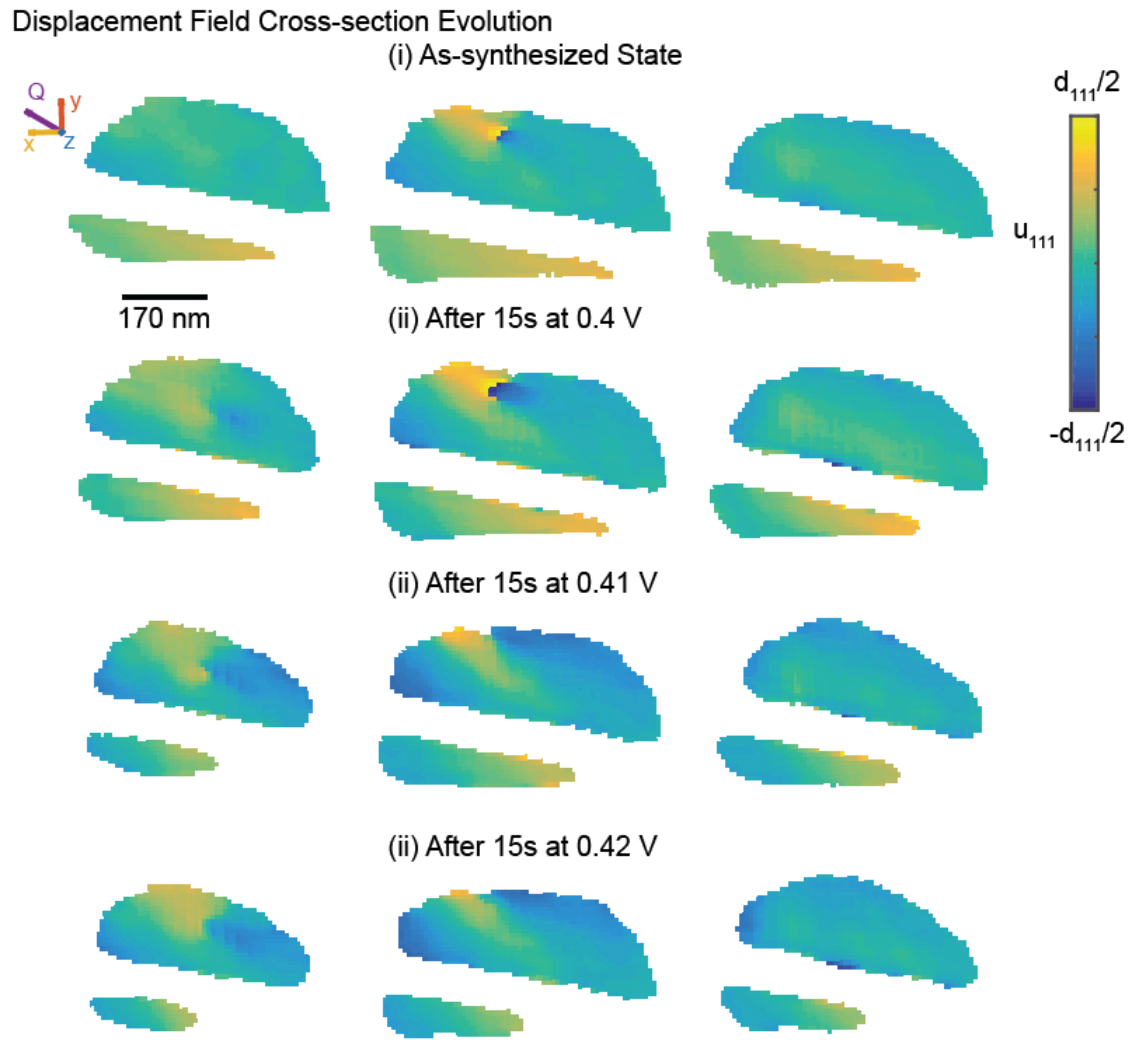

Figure S8. Inhomogeneous dissolution observed in another silver nanocrystal that was not discussed in the main text. Three cross-sections through the particle are shown. The initial state has a pair of dislocations. Over the course of the experiment, the volume of the nanocrystal containing these dislocations dissolved away.

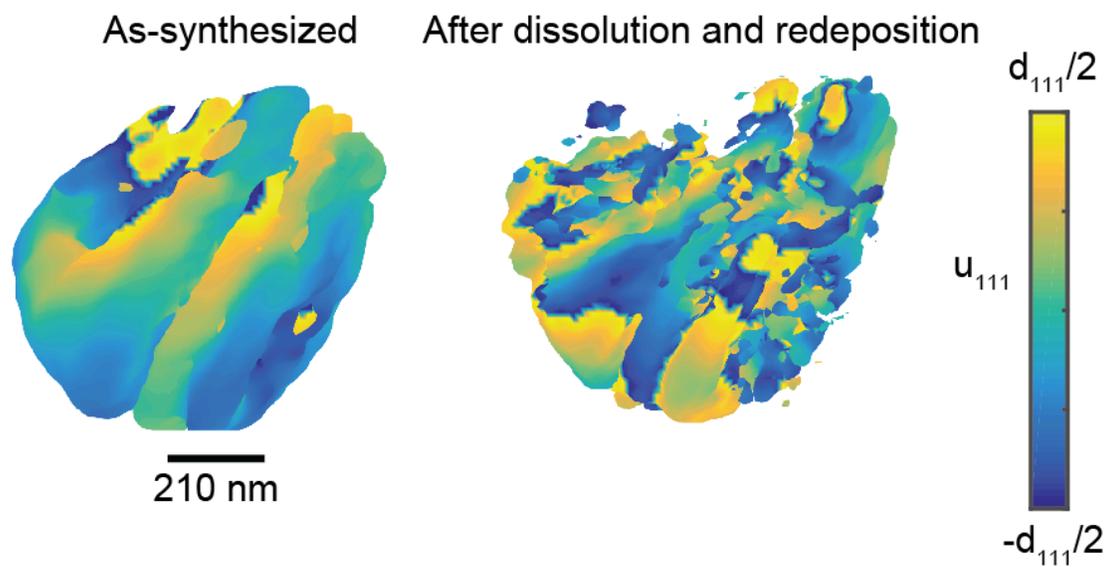

Figure S9. An example of an additional particle that underwent dissolution with redeposition. This particle was not presented in the main text.